\documentclass[conference]{IEEEtran}
\makeatletter
\def\ps@headings{%
\def\@oddhead{\mbox{}\scriptsize\rightmark \hfil \thepage}%
\def\@evenhead{\scriptsize\thepage \hfil \leftmark\mbox{}}%
\def\@oddfoot{}%
\def\@evenfoot{}}
\makeatother
\usepackage{graphicx}
\usepackage{times}
\usepackage{footnote}

\usepackage{graphics}
\usepackage{makecell}
\usepackage{url}
\usepackage{subfig}
\usepackage{epsfig}
\usepackage{caption}
\usepackage{varwidth}
\usepackage{algorithm}
\usepackage{algorithmicx,algpseudocode}

\usepackage{mwe} 

\newcommand{\CASE}[1]{\STATE \textbf{case} #1\textbf{:} \begin{ALC@g}}
\newcommand{\ENDCASE}{\end{ALC@g}}

\newcommand{\DEFAULT}{\STATE \textbf{default:} \begin{ALC@g}}
\newcommand{\ENDDEFAULT}{\end{ALC@g}}
\newcommand{\DEFAULTLINE}[1]{\STATE \textbf{default:} }
\usepackage{color}
\usepackage{multirow}
\usepackage{multicol}
\usepackage{hhline}
\usepackage{amsthm}
\usepackage{comment}
\usepackage{tabu}
\usepackage{amsmath,amsfonts,amssymb,amsthm,epsfig,epstopdf,url,array}

\theoremstyle{plain}

\theoremstyle{definition}

\theoremstyle{remark}

\IEEEoverridecommandlockouts
\begin{document}

\title{GRASP: a GReen energy Aware SDN  Platform}
\author{
Garegin Grigoryan \\ grigorg@clarkson.edu  \\Clarkson University
\and
Keivan Bahmani \\ bahmank@clarkson.edu  \\Clarkson University \and
Grayson Schermerhorn \\ schermgr@clarkson.edu  \\Clarkson University
\and
Yaoqing Liu \\liu@clarkson.edu \\
Clarkson University

}
\date{}
\maketitle

\pagestyle{empty}

\begin{abstract}
The transition to renewable energy sources for data centers has become a popular trend in the IT industry. However, the volatility of renewable energy, such as solar and wind power, impedes the operation of green data centers. In this work, we leverage Software Defined Networking (SDN) to build GRASP, a platform that schedules job requests to distributed data centers according to the amount of green energy available at each site. GRASP can be re-configured with different scheduling algorithms to address diverse factors such as amounts of instantly available solar power, wind power and CPU load of data centers. We utilize realistic green energy datasets from National Solar Radiation Database and evaluate GRASP in the GENI testbed; in addition, we create necessary GENI artifacts to repeat our experiment. GRASP can serve as a practical platform to test various job scheduling mechanisms for distributed green data centers. 
\end{abstract}
\section{Introduction}
\label{sec:intro}
The IT industry has entered the era of Big Data. Efficient and fast Big Data processing is critical for innovations in many areas such as science, healthcare, commerce. The increased demand for high computational power and large storage spaces has inspired the development of an assortment of cloud services, which are being hosted by the growing number of power hungry data centers. The results presented in the US Data Center Energy Usage Report~\cite{shehabi2016united} predicted that in 2020 data centers will consume around 73 billion kWh. Although the report shows the decline of energy demand growth rates due to more efficient design of data centers and the hardware, the leaders of tech industry such as Amazon, Facebook and Microsoft are pursuing data centers fully supplied by green energy~\cite{greendatacenter}. Apple completed this goal for its data centers throughout the world in 2013~\cite{apple}. In 2017, Google announced the plans for purchasing renewable energy in order to match the demands of its data centers and offices~\cite{google}. However, Google also noted that the full transition towards green energy is complicated for several reasons. Namely, suitable locations for generating green energy are mainly far from potential users~\cite{google}. Additionally, the amount of solar and wind power varies throughout the day, seasons and years~\cite{nsrdb}. For instance, solar energy is not available at night or is limited during rainy or cloudy days. On the other hand, data centers are expected to operate continuously and thus consume energy perpetually. Furthermore, storing excess electricity is undesirable for data centers due to its difficulty and cost~\cite{jossen2004operation}. 

In this work, we address the challenge of  instantaneous utilization of green energy in data centers by leveraging Software Defined Networking (SDN) technologies. We introduce \textit{\underline{Gr}}een Energy \textit{\underline{A}}ware \textit{\underline{S}}DN \textit{\underline{P}}latform (GRASP), a platform that balances the incoming jobs to distributed data centers according to various factors, such as the amount of available green energy and the instantaneous computing load of each data center. 

SDN is a new networking technology that decouples the control plane from the data plane. More specifically, the control plane, i.e. SDN controller, is responsible for managing the rules in the forwarding table of the data plane. The data plane, e.g. OpenFlow~\cite{mckeown2008openflow} switch is responsible for matching packets' headers against the forwarding table and select the next hop accordingly. A single controller can control multiple switches remotely when necessary. In SDN networks, the end users or applications can deliver a wide spectrum of useful information to the controller. The controller can further make decisions such as traffic redirection, traffic policing and program the switches based on the information collected from the network. 

The OpenFlow protocol allows the control plane to access the data plane and modify its forwarding rules. In addition, the data plane can send messages to the control plane in the following cases: (1) a switch connects to the controller; (2) a packet matched an entry in the forwarding table with ``send to the controller" action; (3) an entry either expired or removed. In this work we use Ryu SDN framework~\cite{ryu}, that provides a well-defined API for the controller to communicate with one or more OpenFlow switches. However, GRASP can be easily ported to other SDN frameworks as well.


Our contributions in this work are as follows:

	(1) We designed GRASP to effectively load balance jobs among multiple data centers based on the instantaneous amount of green energy available.  The platform can be easily extended to consider additional factors for load balancing and cost-saving purposes. 
	
	(2) We introduced an exemplar green energy-aware algorithm to maximize the utilization of green energy among many distributed data centers. Note, that our aim was not to design a sophisticated scheduling algorithm, but to demonstrate the operation of GRASP only. The design of GRASP is flexible enough to port other more complicated algorithms quickly. 
	
	(3) We implemented and deployed a prototype of GRASP on the GENI testbed~\cite{whatisgeni} with a Ryu SDN controller and multiple OpenFlow switches. We further developed a series of tools that can facilitate researchers to repeat the experiments and reproduce the results.
	
	(4) In our experiment, we utilized realistic green-energy data from National Solar Radiation Database (NSRDB)~\cite{nsrdb}. The results demonstrate the effectiveness of GRASP platform as well as the superior performance of our sample job scheduling algorithm in comparison with the round-robin scheduling.

%

\section{Related Work}
\label{sec:related}
\subsection{Load balancing in data centers}
Load balancing is an important area of research driven by motivation to improve the quality of service in data centers and decrease the operational costs as well. Smart load balancing allows data centers to utilize the available resources in a more efficient manner, avoid links congestion and quickly serve client requests. One of the most commonly used techniques for load balancing is equal-cost multi-path (ECMP) routing~\cite{hopps2000analysis}, when the hash of a packet's 5-tuple determines its next hop. Digit-Reversal Balancing (DRB)~\cite{cao2013per} algorithm forwards packet into different directions in the round robin fashion. However, both algorithms do not consider the current state of the network.  CONGA~\cite{alizadeh2014conga}, an algorithm for distributed congestion-aware load balancing, requires switches to share the link congestion information by piggybacking it to the outcoming packets. Eventually, each switch is able to select the best path for a packet in the data center network. CONGA is a hardware-dependent solution and requires large memory resources on a switch, unlike HULA (Hop-by-hop Utilization-aware Load balancing Architecture)~\cite{katta2016hula}, that is applicable to the programmable switches.  

\subsection{Green computing}
The electricity bills for power and cooling infrastructure may constitute even more than 30\% of all data center operational expenses (\cite{criticalfactors, ibmdatacenter}). This fact as well as the increased carbon emissions by data centers motivates academia and industry to find solutions for (a) Optimizing power consumption in data centers; (b) Using environment-friendly renewable energy sources for supplying data centers with no negative impact on performance.

Shang et al. propose a heuristic routing algorithm~\cite{shang2010energy}, that aggregates the data center traffic into the paths with the least possible number of switches. Li et al.  in~\cite{li2014software} introduce software defined green data center network with \textit{exclusive routing}, when every link is occupied by at most one flow. \textit{Willow}~\cite{kant2011willow} migrates the traffic within a single data center according to energy and thermal profiles of its individual components. \textit{Greenslot}~\cite{goiri2011greenslot}, proposed by Giori et al., schedules jobs within a single data center to maximize the utilization of the renewable energy. The scheduling is planned based on the deadlines of each job and the predictions on the availability of the solar/wind energy. \textit{GreenWare}~\cite{zhang2011greenware}, proposed by Zhang et al.  dynamically dispatches requests in a distributed cloud data center in order to maximally use the renewable energy within a certain cost budget. \textit{GreenWare} uses an algorithm based on a linear-fractional programming. The redirection in \textit{GreenWare} is implemented using the traditional routing mechanisms deployed in the cloud-scale data centers. In~\cite{huang2016green}, authors propose \textit{Green DataPath}, an SDN platform that routes the traffic in a way that minimizes the energy consumption of TCAM chips in the switches of a network. 

To the best of our knowledge, GRASP is the first platform that leverages Software Defined Networks for dynamic redirection of job requests to a distributed green data center. Although GRASP is specially designed for green computing, it can work in other contexts as well. In our simulation, GRASP's SDN controller organizes flows in the network according to how much available solar energy resources data centers have. The list of the parameters that the GRASP's controller considers for decision making can be re-configured, along with parameters' weights; in addition, the scheduling algorithm can be fully replaced based on the requirements set by the operators of the distributed data center.
 \section{Design}
 \label{sec:design}
 
 \subsection{GRASP overview}
 \begin{figure}
 	\centering
 	\includegraphics[width=3in]{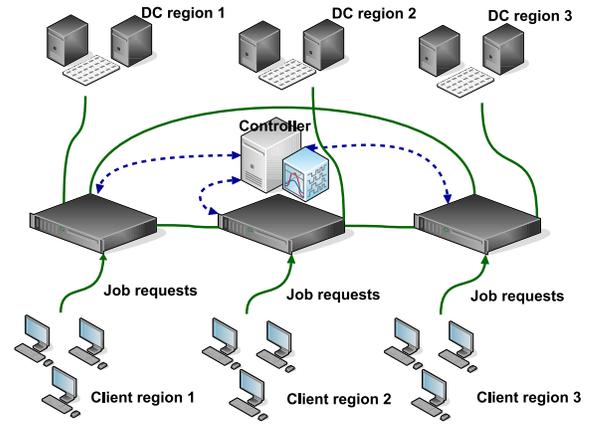}
 	\caption{GRASP overview}
 	\label{fig:framework}
 	\vspace{-6mm}
 \end{figure}
 
 Figure~\ref{fig:framework} illustrates an overview of our GRASP computing platform. Each data center has both green and brown energy available to use. For simplicity, each data center is connected with only one OpenFlow switch. The switches are interconnected together and controlled by a single SDN controller. Clients from each region are connected to their local OpenFlow switch. The requests from clients are directed to those switches. At each OpenFlow switch, the traffic will be re-directed based on the decision made by the controller according to global availability of green energy. The main goal of our current model is to demonstrate the effectiveness of our platform to handle job requests and make forwarding decisions. Hence, in this work we only focus on availability of the green energy. However, more sophisticated scheduling algorithms can be easily ported into our platform to consider other factors, such as network delay or CPU load.
 

 The controller manages the network by installing different forwarding rules into the OpenFlow switches. In other words, the controller can work as a scheduler that makes decisions on which data center will serve the job requests from clients. A straightforward way is to assign a request to the closest data center. However, if the load is high and there is not enough green energy in the closest data center, re-routing job requests to other regions may help improve the quality of service or save costs. In the context of the distributed green data centers, the amount of renewable energy accumulated at a particular data center may vary depending on several factors such as current weather and time of a day or a month. However, the probability of having multiple regions without enough green energy at the same time would be relatively low. For example, when solar panels in Region 1 (see Figure~\ref{fig:framework}) cannot generate green energy due the weather, the data center in Region 3 might possess vast amounts of solar power. In our model, each data center reports their predicted amount of green energy for the next hour to the SDN controller periodically. Based on this data, the controller can make smart decisions to maximize the utilization of green energy at data centers. 
 
 \algnewcommand\algorithmicforeach{\textbf{for each}}
 \algdef{S}[FOR]{ForEach}[1]{\algorithmicforeach\ #1\ \algorithmicdo}
 \algnewcommand\algorithmicswitch{\textbf{switch}}
 \algnewcommand\algorithmiccase{\textbf{case}}
 \algnewcommand\algorithmicassert{\texttt{assert}}
 \algnewcommand\Assert[1]{\State \algorithmicassert(#1)}%
 \algdef{SE}[SWITCH]{Switch}{EndSwitch}[1]{\algorithmicswitch\ #1\ \algorithmicdo}{\algorithmicend\ \algorithmicswitch}%
 \algdef{SE}[CASE]{Case}{EndCase}[1]{\algorithmiccase\ #1}{\algorithmicend\ \algorithmiccase}%
 \algtext*{EndSwitch}%
 \algtext*{EndCase}%
 \begin{algorithm}
 	\caption{Controller's program pseudocode}\label{alg:controller}
 	\begin{algorithmic}[1]
 		\Procedure{$event\_on\_switch\_connect$}{$s$} \Comment{Triggered when the switch $s$ connects to the controller}
 		\State Install the table-miss flow with the lowest priority
 		\State Broadcast $switch\_discover$ packet for establishing the underlying topology
 		\EndProcedure
 		\Procedure{$event\_on\_ip\_packet$}{$p$, $s$}\Comment{Triggered by a miss-flow matched packet $p$ on the switch $s$}
 		\Switch{packet $p$ from the port $i_s$}
 		\Case {$p$ registers a data center $d$:}
 		\State Store $d$'s IP address $d.ip = p.ip\_src$
 		\State Store egress port $d.port = i_s$.
 		\State Store next hop MAC address $d.mac = p.mac\_src$
 		\State Store the id of $s$ connected to $d$ into $d.s$
 		\State Assign id to $d$ and store it into $d.id$
 		\State Generate a passcode for $d.id$
 		\State Send $d.id$, the passcode  and the list of parameters to $p.ip\_src$ through the port $i_s$
 		\EndCase
 		\Case {$p$ is a $switch\_discover$ packet from neighbor switch $s_n$ from the port $i_s$}
 		\State Store next hop MAC address for $s$ to reach $s_n$ $(s, s_n).mac = p.mac\_src$
 		\State Store egress port for $s$ to reach $s_n$ $(s, s_n).port=i_s$
 		\EndCase
 		\Case {$p$ is a packet with parameters' values from a registered $d$:}
 		\State Store parameters' values into $E[d.id]$
 		\EndCase
 		\Case {default:}
 		\State $d \gets desicion\_making$ \Comment{Choosing the best data center $d$ to forward a job request}
 		\State Rewrite $p$'s Ethernet frame and IP layer headers for the next hop of $d$
 		\State Install the flow rules for $p.ip\_src$ in the switch(-es) along the path from the port $i_s$ to $d.port$
 		\If {$d.s \neq s$}
 		\State Forward $p$ to $(s, d.s).port$
 		\Else
 		\State Forward $p$ to the port $d.port$
 		\EndIf
 		\EndCase
 		\EndSwitch
 		\EndProcedure
 		
 	\end{algorithmic}
 \end{algorithm}
 
 \begin{algorithm}
 	\caption{Green energy-aware scheduler's pseudocode}\label{alg:dc}
 	\begin{algorithmic}[1]
 		\Procedure{$decision\_making$}{}\Comment $k$ - amount of the energy consumed by a job; $E[d.id]$ - amount of the green energy available at the data center $d$; $n[d.id]$ - the number of active jobs at the data center $d$.
 		\ForEach {data center $d$ in the network}
 		\State $G[d.id]=\frac{E[d.id]}{k} - n[d.id]$
 		\EndFor
 		\State Select $d_s$ such as $G[d_s.id] = max(G)$
 		\State $n[d_s.id] = n[d_s.id] + 1$
 		\State \Return $d_s$
 		\EndProcedure
 	\end{algorithmic}
 \end{algorithm}
 
 \subsection{GRASP Design}
 
\subsubsection{Configuration} Initially, GRASP requires configuration of the controller and agents in each data center. At the controller end, a configuration file is used to record: (1) Required parameters that individual data centers need to send, e.g., amount of green energy, CPU load, memory usage etc.; (2) The weight of each parameter for scheduling decisions; and (3) How often the data centers should report their on-site information to the controller. Based on the configuration file, agents in each data center are required to report different parameters and their values in a fixed format that can be understood and parsed by the controller. 

\subsubsection{Initialization} Before entering into the operation mode, GRASP goes into initialization mode. Initially, for security reasons, each data center should be registered at the SDN controller and be given a key. This key is verified every time a data center informs the controller about it's current state. The registration packets are using the table-miss flow, that is installed in the forwarding table of each switch once they connect to the controller. A packet that matches the table-miss flow is sent to the controller along the switch's ID and the packet's ingress port. Second, the controller stores the IP address of the data center's agent, next hop MAC address, and the port number that connects that data center to the switch. In response, the controller sends back a packet with the following data: (1) A verification key; (2) A list of parameters that should be reported to the controller; (3) The frequency of those messages. Once a data center receives the controller's response, it begins sending the reporting messages with current state of the data center to the controller. Next, the controller considers the data center as a potential candidate to process job requests from end users. Finally, the controller needs to fully comprehend the underlying topology formed by the interconnected OpenFlow switches. To this end, the controller generates a discovery packet with a special key and broadcasts it to each switch using the \textit{OFPP\_FLOOD} OpenFlow command. These packets will match the table-miss flow and be bounced back to the controller along with their ingress switch port numbers. The discovery process allows the controller to be fully aware of the network topology and control the flows in the network, i. e. schedule packets from one region to another.

\subsubsection{Operation} After the initialization, GRASP is ready to process job requests and forward them to the appropriate data centers. Suppose a client from region 1 wants to establish a TCP connection with one of the distributed data centers to send job requests. The first packet of the flow (e.g. TCP SYN packet) will match against the table-miss flow, which will guide the packet to the controller. Once the controller recognizes this packet as a client's request, it runs a decision making algorithm with a scheduler to select the best data center to serve the job request. Based on the chosen data center, the controller takes the following actions: (1) Rewriting the Ethernet frame and IP headers of the packet; (2) Installing all necessary flows into the forwarding tables over the switches on the path between the client and the specified data center; (3) Forwarding the packet of the client to the designated data center. The following data packets coming from the client will match the flows installed in the switches and will not be redirected to the controller. The process indicates that only the first new packet from a client will be handled by the controller, and other data packets will be directly forwarded on the fast data path. This design can significantly enhance the system performance. In GRASP, the flows are installed with a certain $timeout$ value. The value can be set equal to the maximum Round Trip Time (RTT) in the network, so if during $timeout$ no packet matches the flow entries they will be removed from the forwarding tables. Therefore, the controller is not required to redirect packets belonging to existing active TCP connections and makes decisions only for new client requests.

\subsection{Green Energy Aware Scheduler}

To evaluate GRASP, we introduce a simple green energy aware scheduler that works based on several assumptions: (1) The energy consumption in a data center is proportional to the number of jobs being processed; (2) Every job consumes the same amount of energy; (3) An agent from each data center sends to the controller the amount of estimated green energy of the next hour in an hourly basis; (4) Each scheduled job completes within one hour. Based on these assumptions, we made a few changes to the traditional round-robin scheduling algorithm. Intuitively, a new job will be assigned to whichever data center that has the largest amount of available green energy at that time. Once a job was assigned to a data center, the total available green energy of the data center will be deducted by the units of energy consumed by the job. Therefore, the outcome of the scheduling algorithm depends on the amount of green energy generated at each data center and the number of active jobs being processed. This way, the data centers will take turn to run jobs if they have similar amounts of green energy. Thus, our algorithm can be called green energy aware round-robin scheduling algorithm. 

Assume that $d$ is a data center and $G_d$ is the number of jobs the data center $d$ can serve with its green energy. Then:
\begin{equation*}
\label{eq:gd}
G_d = \frac{E_d}{k} - n_d,
\end{equation*}
where (a) $E_d$ is the amount of green energy that was available at time $t$ in a data center $d$, (b) $k$ is the energy required to process a single job, and (c) $n_d$ is the number of active jobs already scheduled to the data center $d$. For a new job request, the scheduler selects a data center $d_s$, where $G_{d_s} = max(G_{d_1}, G_{d_2}, ..., G_{d_m})$ and $m$ is the total number of the green data centers. We show the pseudocode of the main components of GRASP and the green energy-aware scheduler in Algorithms~\ref{alg:controller} and~\ref{alg:dc}. We implemented the prototype using Ryu SDN controller~\cite{ryu}. In Section~\ref{sec:evaluation}, we evaluated the performance of our new scheduler VS the traditional round-robin scheduler, with which each data center takes turn to process each job request. Note that the primary goal of this paper is NOT to propose an optimal scheduling algorithm to minimize costs, but to showcase the effectiveness and performance of the GRASP platform. Operators of the distributed data centers can consider many other parameters to tune or even entirely replace our algorithm to meet their requirements.

\section{Evaluation}
\label{sec:evaluation}
\begin{table}[tbp]
	\centering
	\begin{tabular}{|l|l|l|}
		\hline 
		\textbf{New York} & \textbf{Florida} & \textbf{California} \\ \hline 
		1. Elmira Corning Regional & 4. Homestead & 7. Lompoc \\ \hline
		2. Watertown & 5. Orlando & 8. March\\ \hline
		3. Westhampton Gabreski & 6. Tyndall & 9. Travis Field\\ \hline
	\end{tabular}
	\captionsetup{justification=centering}
	\caption{The list of data centers at different places}
	\label{tab:locations}
	\vspace{-1mm}
\end{table}
\begin{figure}[t]
	\centering
	\includegraphics[width=2.2in]{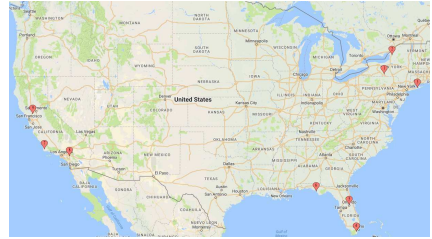}
	\vspace{-1mm}	
	\captionsetup{justification=centering, width=\linewidth}
	\caption{The map of the simulated data centers}
	\label{fig:usmap}
		\vspace{-6mm}
\end{figure}
\begin{figure*}[t]
	\centering
	\includegraphics[width=\linewidth]{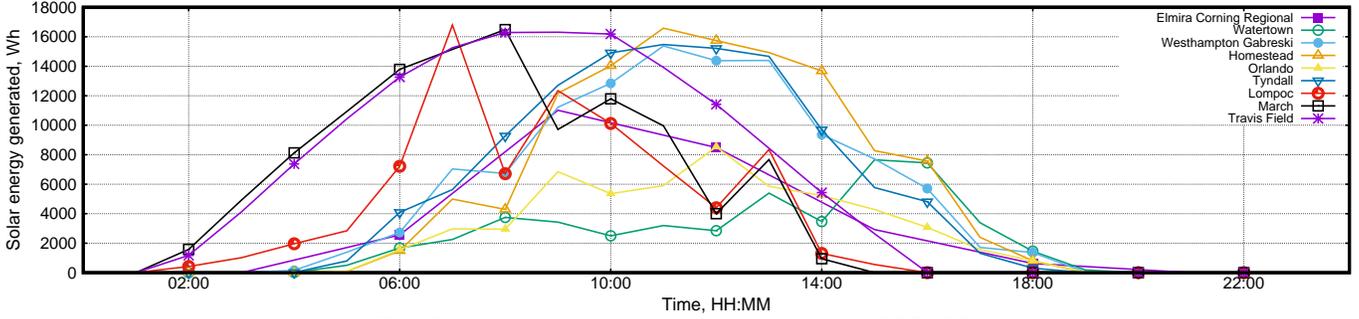}
	\vspace{-6mm}	
	\captionsetup{justification=centering, width=\linewidth}
	\caption{Amounts of solar energy available on 06/21/2001}
	\label{fig:energy}
		\vspace{-5mm}
\end{figure*}
\begin{figure*}[t]
	\centering
	\includegraphics[width=\linewidth]{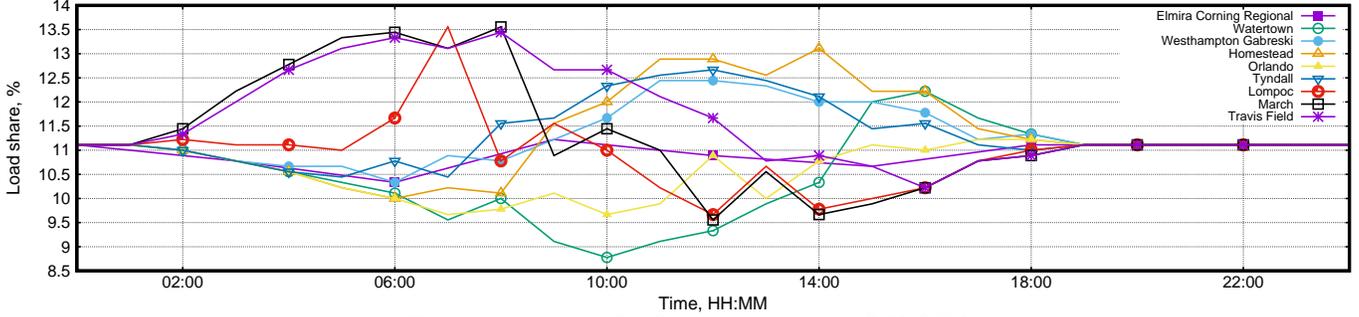}
	\vspace{-6mm}	
	\captionsetup{justification=centering, width=\linewidth}
	
	\caption{Load share for 9 data centers on 06/21/2001}
		\vspace{-6mm}
	\label{fig:load}
\end{figure*}

To simulate the distributed data center operation, we utilized the data from National Solar Radiation Database (NSRDB)~\cite{nsrdb}. This data was retrieved from data gathering stations located around the United States and captured every hour during an array of years. The data has been combined into a simulated single year based on the most accurate data. In our evaluation we selected 9 locations in New York, Florida, and California (see Table~\ref{tab:locations} and Figure~\ref{fig:usmap}). Using the data pertaining to the Dry Bulb Temperature (Celsius) and the Global Horizontal Radiation (Wh/m2), we were able to generate the full approximation for the solar power generation (Wh) for every hour of every day in a year at each of these locations. Our experiment simulated a scenario with 9 green data centers that are provided with the solar energy from above-mentioned locations. Each hour, the controller that manages and controls each of those green data centers gets a certain number of job requests from its clients. The goal of GRASP is to balance these requests in order to maximize the total usage of solar energy.

\begin{figure}[t]
	\centering
	\includegraphics[width=0.7\linewidth]{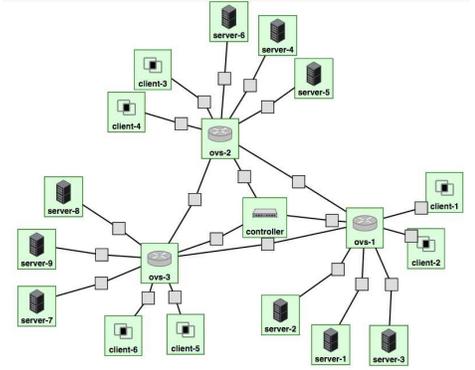}
	\vspace{-2mm}	
	\captionsetup{justification=centering, width=\linewidth}
	\caption{GENI topology}
	\label{fig:genifig}
		\vspace{-6mm}
\end{figure}

 We implemented and deployed GRASP in GENI testbed~\cite{whatisgeni} with 9 servers simulating 9 data centers, 3 interconnected OpenFlow switches and 6 clients generating web requests to verify it's performance. Figure~\ref{fig:genifig} illustrates the GENI topology used in our experiment, while   Figure~\ref{fig:energy} illustrates the values of the solar energy generated during June 21st (the summer solstice day) reported by 9 data centers to the controller. It can be observed from Figure~\ref{fig:load} that GRASP's green energy-aware scheduler assigned jobs proportionally between data centers and minimized the number of scheduled jobs to the data centers with less available solar energy.  Note that, if there is no solar energy available at all data centers, our scheduling algorithm works as a simple round-robin scheduler, giving each of the 9 data centers $\approx$11.1\% of the total computing load. 

In the rest of this section we compare the performance of the green energy-aware scheduler with the traditional round robin scheduler, when each data center gets the same load independent of its solar energy reserves. Because our goal was to compare two schedulers given the different load and energy consumption values over a year long period, we sped up the experiment by running the simulation on a single host rather than using GENI resources. For both schedulers, we measured the hourly ratio $r$ between the number of jobs served by the green energy $n_{g}$ and the total number of processed jobs $n$. Obviously, in all cases, $n_{g}\leq n$. Furthermore, $n_{g}$=$n$, i. e. $r$=1 can be only achieved if data centers have enough green energy to process all the jobs. We calculated $n_{g}$ for each simulated hour of distributed data center's operation as follows:
\begin{enumerate}
\item For each data center $d$ we calculate $max_d = \frac{E_d}{k}$, where $max_d$ is the maximum number of jobs that can be executed with green energy in $d$ during an hour, $E_d$ is the available green energy in $d$ during that hour, and $k$ is the energy required for processing a single job with unit Wh.
\item $n_{g} = \displaystyle\sum_{Each~d}min(max_d, n_d)$, where $n_d$ is the actual number of jobs processed in $d$ during that hour. Note, that $n_d$ may be larger than $max_d$ when there is not enough green energy to process all the jobs scheduled to $d$. 
\end{enumerate} 
The value of $k$ may vary depending on what type of job a data center processes. We assumed the same value of $k$ for each job processed by any of the green data centers in our model and we ran experiments with $k$ ranging from 1 to 200 and 900 jobs per each hour. For each $k$, we compare the 1-year average ratio for $r$, $r_{avg}$, achieved by the green energy-aware scheduler and the round-robin scheduler. The bigger values of $k$, the narrower is the performance gap between the two algorithms, because when $k$ is larger, the total energy demand may far exceed the total green energy supply for all data centers.

\begin{figure}[t]
	\centering
	\includegraphics[width=0.7\linewidth]{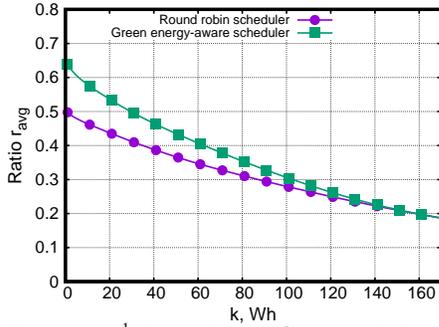}
	\vspace{-2mm}	
	\captionsetup{justification=centering, width=\linewidth}
	\caption[Caption for LOF]{Ratio $r_{avg}$\footnotemark \space for different $k$ given 900 jobs per hour}
	\label{fig:ratio}
	\vspace{-4mm}
\end{figure}

The results for a 1-year average ratio $r_{avg}$ between jobs handled by green energy and total jobs are illustrated in Figure~\ref{fig:ratio}. For minimal values of $k$, our green energy-aware scheduler achieves better $r_{avg}$ by nearly 15\%. However, the difference between the performance of two schedulers decreases with increased $k$. We assume that more diverse patterns of renewable energy generation, e.g, wind power, may enlarge the performance difference between two schedulers. Also GRASP can be configured with other schedulers that consider more factors, e.g., CPU load, to improve the overall system performance.  

\begin{figure}[t]
	\centering
	\includegraphics[width=0.7\linewidth]{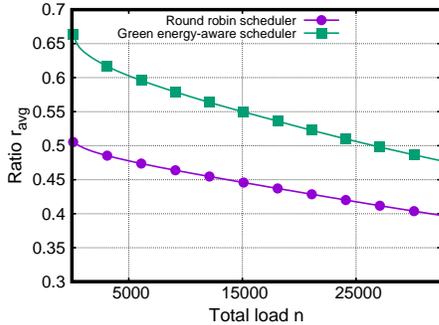}
	\vspace{-2mm}	
	\captionsetup{justification=centering, width=\linewidth}
	\caption{Ratio $r_{avg}$ for different job requests load per hour when $k$=1Wh}
	\label{fig:ratio2}
	\vspace{-8mm}
\end{figure}

Figure~\ref{fig:ratio2} illustrates how the 1-year average ratio $r_{avg}$ changes when we set $k$=1Wh and change the number of total job requests per hour. The results show that given the minimum load, our green energy-aware scheduler uses 16\% more solar energy on average than round-robin scheduler. We also observed that increasing the number of job requests leads to decrease in $r_{avg}$ of both schedulers since more jobs consume more brown energy on the grid. In summary, our GRASP platform can be a valuable tool to test and evaluate the performance of various scheduling algorithms for a distributed green data center with different settings in terms of diverse job requests and green energy generation patterns. 
 \footnotetext{The 1-year average ratio between the number of jobs served by green energy and the total number of jobs.}

\section{Conclusion}
\label{sec:conclusion}
In the era of distributed computing data centers,  operators seek to reduce their negative impact on the environment and optimize their energy utilization by moving towards renewable energy. In this paper, we introduce a Green Energy Aware SDN Computing Platform \textit{GRASP} for scheduling jobs to data centers in order to maximize the utilization or the renewable energy. Our platform is based on Software Defined Networking with a centralized controller that collects the information from data centers in different regions and schedules jobs for them. GRASP can be used for operating green storages, mining Bitcoins in a distributed green farm. Additionally, we illustrated the effectiveness of GRASP using two job schedulers through the deployment on the GENI tesbed and simulating the operation of a distributed green data center in a 1-year period using realistic data.

\bibliographystyle{IEEEtran}
\bibliography{net}
\end{document}